\input harvmac
\noblackbox
%%% Figures
\newcount\figno
\figno=0
\def\fig#1#2#3{
\par\begingroup\parindent=0pt\leftskip=1cm\rightskip=1cm\parindent=0pt
\baselineskip=11pt
\global\advance\figno by 1
\midinsert
\epsfxsize=#3
\centerline{\epsfbox{#2}}
\vskip 12pt
\centerline{{\bf Figure \the\figno:} #1}\par
\endinsert\endgroup\par}
\def\figlabel#1{\xdef#1{\the\figno}}

\def\np#1#2#3{Nucl. Phys. {\bf B#1} (#2) #3}
\def\pl#1#2#3{Phys. Lett. {\bf B#1} (#2) #3}
\def\prl#1#2#3{Phys. Rev. Lett.{\bf #1} (#2) #3}
\def\physrev#1#2#3{Phys. Rev. {\bf D#1} (#2) #3}

%%% Paragraphs

%%% special math symbols
\font\cmss=cmss10
\font\cmsss=cmss10 at 7pt
\def\rlx{\relax\leavevmode}
\def\inbar{\vrule height1.5ex width.4pt depth0pt}
\def\IC{\relax\,\hbox{$\inbar\kern-.3em{\rm C}$}}
\def\IN{\relax{\rm I\kern-.18em N}}
\def\IP{\relax{\rm I\kern-.18em P}}
\def\ZZ{\rlx\leavevmode\ifmmode\mathchoice{\hbox{\cmss Z\kern-.4em Z}}
 {\hbox{\cmss Z\kern-.4em Z}}{\lower.9pt\hbox{\cmsss Z\kern-.36em Z}}
 {\lower1.2pt\hbox{\cmsss Z\kern-.36em Z}}\else{\cmss Z\kern-.4em
 Z}\fi} 
%%% misc.
\def\IZ{\relax\ifmmode\mathchoice
{\hbox{\cmss Z\kern-.4em Z}}{\hbox{\cmss Z\kern-.4em Z}}
{\lower.9pt\hbox{\cmsss Z\kern-.4em Z}}
{\lower1.2pt\hbox{\cmsss Z\kern-.4em Z}}\else{\cmss Z\kern-.4em
Z}\fi}

\def\narrowplus{\kern -.04truein + \kern -.03truein}
\def\narrowminus{- \kern -.04truein}
\def\narrowminussub{\kern -.02truein - \kern -.01truein}

\def\kh{K\"{a}hler }

\def\r{{\rightarrow}}

\def\frac#1#2{{#1\over #2}}

\def\CM{{\cal M}}

\def\BR{\IR}

\def\IZ{\relax\ifmmode\mathchoice
{\hbox{\cmss Z\kern-.4em Z}}{\hbox{\cmss Z\kern-.4em Z}}
{\lower.9pt\hbox{\cmsss Z\kern-.4em Z}}
{\lower1.2pt\hbox{\cmsss Z\kern-.4em Z}}\else{\cmss Z\kern-.4em
Z}\fi}
\def\IB{\relax{\rm I\kern-.18em B}}
\def\IC{{\relax\hbox{$\inbar\kern-.3em{\rm C}$}}}
\def\ID{\relax{\rm I\kern-.18em D}}
\def\IE{\relax{\rm I\kern-.18em E}}
\def\IF{\relax{\rm I\kern-.18em F}}
\def\IG{\relax\hbox{$\inbar\kern-.3em{\rm G}$}}
\def\IGa{\relax\hbox{${\rm I}\kern-.18em\Gamma$}}
\def\IH{\relax{\rm I\kern-.18em H}}
\def\II{\relax{\rm I\kern-.18em I}}
\def\IK{\relax{\rm I\kern-.18em K}}
\def\IP{\relax{\rm I\kern-.18em P}}
%\def\IX{\relax{\rm X\kern-.01em X}}
%this doesn't work

\font\cmss=cmss10 \font\cmsss=cmss10 at 7pt
\def\IR{\relax{\rm I\kern-.18em R}}

\def\S{{\Sigma}}
%

%
%       \eqn\label{a+b=c}	gives displayed equation, numbered
%				consecutively within sections.
%     \eqnn and \eqna define labels in advance (of eqalign?)
%
\def\eqnn#1{\xdef #1{(\secsym\the\meqno)}\writedef{#1\leftbracket#1}%
\global\advance\meqno by1\wrlabeL#1}
\def\eqna#1{\xdef #1##1{\hbox{$(\secsym\the\meqno##1)$}}
\writedef{#1\numbersign1\leftbracket#1{\numbersign1}}%
\global\advance\meqno by1\wrlabeL{#1$\{\}$}}
\def\eqn#1#2{\xdef #1{(\secsym\the\meqno)}\writedef{#1\leftbracket#1}%
\global\advance\meqno by1$$#2\eqno#1\eqlabeL#1$$}

\lref\rpol{J. Polchinski, ``TASI Lectures on D-Branes,''
hep-th/9611050\semi J. Polchinski, S. Chaudhuri and C. Johnson,
``Notes on D-Branes,'' hep-th/9602052. } 
\lref\rBFSS{T. Banks, W. Fischler, S. H. Shenker, and L. Susskind, ``M
Theory As A Matrix Model: A Conjecture,''  
hep-th/9610043, Phys. Rev. {\bf D55} (1997) 5112.}
\lref\rwtensor{E. Witten, ``Some Comments on String Dynamics,''
hep-th/9507121.}
\lref\rstensor{A. Strominger, ``Open P-Branes,'' hep-th/9512059,
\pl{383}{1996}{44}.} 
\lref\rsdecoupled{N. Seiberg, ``New Theories in Six Dimensions and
Matrix Description of M-theory on $T^5$ and $T^5/\IZ_2$,''
hep-th/9705221.}
\lref\rashoke{ A. Sen, ``Kaluza-Klein Dyons in String
Theory,'' hep-th/9705212; ``A Note on Enhanced Gauge Symmetries in
M and String Theory,'' hep-th/9707123; ``Dynamics of Multiple
Kaluza-Klein Monopoles in M and String Theory,'' hep-th/9707042.}
\lref\kutetal{D. Berenstein, R. Corrado and J. Distler, ``On the
Moduli Spaces of M(atrix)-Theory Compactifications,''
hep-th/9704087\semi  S. Elitzur, A. Giveon, D. Kutasov and
E. Rabinovici, ``Algebraic Aspects of Matrix Theory on $T^d$,''
hep-th/9707217.} 
\lref\rtwoform{J. P. Gauntlett and D. Lowe, ``Dyons and S-Duality in
N=4 Supersymmetric Gauge Theory,'' hep-th/9601085,
\np{472}{1996}{194}\semi K. Lee, E. Weinberg and P. Yi,
``Electromagnetic Duality and $SU(3)$ Monopoles,'' hep-th/9601097,
\pl{376}{1996}{97}.} 
\lref\rmoore{A. Losev, G. Moore, and S. Shatashvili, ``M \& m's ,''
hep-th/9707250.}
\lref\rbrunner{I. Brunner and A. Karch, ``Matrix Description of
M-theory on $T^6$,'' hep-th/9707259.}
\lref\rDVV{R. Dijkgraaf, E. Verlinde and H. Verlinde, ``BPS Spectrum
of the Five-Brane and Black Hole Entropy,'' hep-th/9603126, 
\np{486}{1997}{77}; ``BPS Quantization of the Five-Brane,''
hep-th/9604055, \np{486}{1997}{89}.} 
\lref\rsixbrane{P. Townsend, ``The Eleven Dimensional Supermembrane
Revisited,'' hep-th/9501068, \pl{350}{1995}{184}.} 
\lref\rmultitn{S. Hawking, ``Gravitational Instantons,''
Phys. Lett. {\bf 60A} (1977) 81\semi 
G. Gibbons and S. Hawking, ``Classification of Gravitational Instanton
Symmetries,'' Comm. Math. Phys. {\bf 66} (1979) 291\semi
R. Sorkin, ``Kaluza-Klein Monopole,''
\prl{51}{1983}{87}\semi D. Gross and M. Perry, ``Magnetic Monopoles in
Kaluza-Klein Theories,'' \np{226}{1983}{29}.}
\lref\rmIIB{P. Aspinwall, ``Some Relationships Between Dualities in
String Theory,'' hep-th/9508154, Nucl. Phys. Proc. Suppl. {\bf 46}
(1996) 30\semi J. Schwarz, ``The Power of M Theory,'' 
hep-th/9510086, \pl{367}{1996}{97}. }
\lref\rgeneralsixbrane{ C. Hull,  ``Gravitational Duality, Branes and
Charges,'' hep-th/9705162\semi E. Bergshoeff, B. Janssen, and
T. Ortin, ``Kaluza-Klein Monopoles and Gauged Sigma Models,''
hep-th/9706117\semi Y. Imamura, ``Born-Infeld Action and Chern-Simons 
Term {}from Kaluza-Klein Monopole in M-theory,'' hep-th/9706144.}
\lref\rbd{M. Berkooz and M. Douglas, ``Five-branes in M(atrix)
Theory,'' hep-th/9610236, \pl{395}{1997}{196}.}
\lref\rbraneswith{M. Douglas, ``Branes within Branes,''
hep-th/9512077.} 
\lref\rquantumfive{O. Aharony, M. Berkooz, S. Kachru, N. Seiberg, and
E. Silverstein,`` Matrix Description of Interacting Theories in Six
Dimensions,'' hep-th/9707079.}
\lref\rstringfive{E. Witten, ``On The Conformal Field Theory of The
Higgs Branch,'' hep-th/9707093.}
\lref\rSS{S. Sethi and L. Susskind, ``Rotational Invariance in the
M(atrix) Formulation of Type IIB Theory,'' hep-th/9702101, 
\pl{400}{1997}{265}.} 
\lref\rBS{T. Banks and N. Seiberg, ``Strings from Matrices,''
hep-th/9702187, \np{497}{1997}{41}.} 
\lref\rgilad{A. Hanany and G. Lifschytz, ``M(atrix) Theory on $T^6$
and a m(atrix) Theory Description of KK Monopoles,'' hep-th/9708037.}
\lref\rreview{N. Seiberg, ``Notes on Theories with 16 Supercharges,''
hep-th/9705117.}
\lref\rDVVstring{R. Dijkgraaf, E. Verlinde and H. Verlinde, ``Matrix
String Theory,'' hep-th/9703030.}
\lref\rprobes{M. Douglas, ``Gauge Fields and D-branes,''
hep-th/9604198.}
\lref\doumoo{M. Douglas and G. Moore, ``D-Branes,
Quivers, and ALE Instantons,'' hep-th/9603167.}
\lref\rfischler{M. Douglas, ``Enhanced Gauge Symmetry in M(atrix)
Theory,'' hep-th/9612126\semi
W. Fischler and A. Rajaraman, ``M(atrix) String Theory
on K3,'' hep-th/9704123\semi
C. Johnson and R. Myers, ``Aspects of Type IIB Theory on ALE Spaces,''
hep-th/9610140, \physrev{55}{1997}{6382}\semi
D.-E. Diaconescu and J. Gomis, ``Duality in Matrix Theory
and Three Dimensional Mirror Symmetry,'' hep-th/9707019.}
\lref\rsprobes{N. Seiberg, ``Gauge Dynamics And Compactification To
Three Dimensions,'' hep-th/9607163, \pl{384}{1996}{81}} 
\lref\rswthree{N. Seiberg and E. Witten, ``Gauge Dynamics and
Compactifications to Three Dimensions,'' hep-th/9607163.}
\lref\rthroat{D.-E. Diaconescu and N. Seiberg, ``The Coulomb Branch of
$(4,4)$ Supersymmetric Field Theories in Two Dimensions,''
hep-th/9707158. }
\lref\rTduality{T. Banks, M. Dine, H. Dykstra and W. Fischler,
``Magnetic Monopole Solutions of String Theory,''
\pl{212}{1988}{45}\semi C. Hull and P. Townsend, ``Unity of
Superstring Dualities,'' hep-th/9410167, \np{438}{109}{1995}\semi
H. Ooguri, C. Vafa, ``Two Dimensional Black Hole and Singularities of
Calabi-Yau Manifolds,'' Nucl.Phys. {\bf B463} (1996) 55, 
hep-th/9511164\semi D. Kutasov, ``Orbifolds and Solitons,'' Phys. Lett
{\bf B383} (1996) 48, hep-th/9512145\semi H. Ooguri and C. Vafa,
``Geometry of N=1 Dualities in Four Dimensions,'' hep-th/9702180.}
\lref\raps{P. Argyres, R. Plesser and N. Seiberg, ``The Moduli Space
of N=2 SUSY QCD and Duality in N=1 SUSY QCD,'' hep-th/9603042,
\np{471}{1996}{159}.} 
\lref\rgms{O. Ganor, D. Morrison and N. Seiberg, ``Branes, Calabi-Yau
Spaces, and Toroidal Compactification of the N=1 Six Dimensional $E_8$
Theory,'' hep-th/9610251, \np{487}{1997}{93}.} 
\lref\rchs{ C.G. Callan, J.A. Harvey, A. Strominger, ``Supersymmetric
String Solitons,'' hep-th/9112030, \np{359}{1991}{611}\semi S.-J. Rey,
in ``The  Proc. of the Tuscaloosa Workshop 
1989,'' 291; Phys. Rev. {\bf D43} (1991) 526; S.-J. Rey, In DPF Conf.
1991, 876.}
\lref\rmotl{L. Motl, ``Proposals on nonperturbative superstring
interactions,'' hep-th/9701025.} 
\lref\rwati{W. Taylor IV, ``D-Brane Field Theory on Compact Space,''
hep-th/9611042, \pl{394}{1997}{283}.} 
\lref\kin{K. Intriligator, to appear.}

\Title{\vbox{\hbox{hep-th/9708085}\hbox{IASSNS-HEP-97/94}}}
{Comments on Neveu-Schwarz Five-Branes}

\smallskip
\centerline{Nathan Seiberg\footnote{$^1$}{seiberg@sns.ias.edu} and
Savdeep Sethi\footnote{$^2$} {sethi@sns.ias.edu} }
\medskip\centerline{\it School of Natural Sciences}
\centerline{\it Institute for Advanced Study}\centerline{\it
Princeton, NJ 08540, USA}

%%
%% Draft of paper defining brane theories with compact transverse
%% dimensions
%%

\vskip 1in

\noindent
We study the theory of NS five-branes in string theory with a smooth
non-trivial transverse space.  We show that in the limit that the bulk
physics decouples, these theories become equivalent to theories with a
flat and non-compact transverse space. We present a matrix model
description of the type IIA theory on $\IR^9\times S^1$ with NS
five-branes located at points on the circle. Consequently, we obtain a
description of the dual configuration of Kaluza-Klein monopoles in the
type IIB theory.

%\draftmode
\vskip 0.1in
\Date{8/97}

\newsec{Introduction}

The five-branes of M theory and string theory are extremely
interesting objects \rDVV.  The theory on $k$ coincident five-branes
in M theory is an interacting field theory at a non-trivial fixed
point of the renormalization group. This theory was first found in
\refs{\rwtensor, \rstensor}; for a review, see e.g.\ \rreview.  We
will refer to this theory as the $(2,0) $ field theory.  To obtain
this field theory, we have to consider the limit where the eleven
dimensional Planck scale $M_{pl}$ goes to infinity.  In this limit,
all interactions with the modes in the bulk of spacetime, including
the interactions with gravity, decouple and we are left with a {\it
complete} theory on the five-branes.  The moduli space of vacua for
this theory is
\eqn\modsf{ \CM = {(\IR^5)^k\over {\bf S}_k}.}
These theories are naturally associated with the $A_{k-1}$ groups.
Extensions to other groups and in particular to $D_k$ and $E_{6,7,8}$
were discussed in \refs{\rwtensor, \rreview}. For other groups, the
quotient by the permutation group is replaced by the appropriate Weyl
group.

A generalization of this theory was found in \rsdecoupled.  There, M
theory on $\IR^{10}\times S^1$ was studied with $k$ five-branes at
points on the circle.  To find a complete theory, we should again make
sure that the modes on the five-branes decouple {}from the modes in
the bulk of spacetime.  Again, this is achieved by considering the
limit $M_{pl} \rightarrow \infty$.  However, unlike the previous
case, we now have another parameter -- the radius of the
circle $L$.  Therefore, we can find a family of new theories which
depend on this parameter.  More specifically, by taking
\eqn\limits{\eqalign{
&M_{pl} \rightarrow \infty, \cr
&L \rightarrow 0, \cr}}
while holding fixed
\eqn\msfi{M_s^2= L M_{pl}^3, }
we find a new theory which depends on $M_s$.  Equivalently, by
starting with the type IIA theory, rather than with M theory, we can
define this theory by taking the string coupling $g_s^A$ to zero,
while holding fixed the string scale $M_s$.  We will refer to this
theory as the $(2,0)$ string theory, since it includes string-like
excitations with tension $M_s^2$.  The moduli space of vacua is now
\eqn\modss{\CM = {(\IR^4 \times S^1)^k\over {\bf S}_k},}
where the radius of the $S^1$ factor is 
\eqn\tzsr{P=LM_{pl}^3=M_s^2.}
This follows since $P$ is clearly proportional to $L$ and the factor
of $M_{pl}^3$ appears  on dimensional grounds, since $M_{pl}$ is the
only scale in the problem.  In the ``zero slope limit,'' $M_s
\rightarrow \infty$, this theory reduces to the $(2,0)$ field theory.
As a check, note that in this limit \modss\ becomes the same as
\modsf.  These theories are naturally associated with the $A_{k-1}$
groups.  Extensions to other groups, and in particular to $D_k$ and
$E_{6,7,8}$, are straightforward.

Another ``non-critical string theory'' with $(1,1)$ supersymmetry is
similarly obtained by starting with $k$ NS five-branes in type IIB
string theory, in the limit where the string coupling $g_s^B$
vanishes with the string scale held fixed \rsdecoupled.  After
compactification on a longitudinal circle of radius $R$, these $(1,1)$
string theories are the same as the $(2,0)$ string theories
compactified on a circle of radius $1 \over RM_s^2$.  This fact has
led to the conclusion that these theories are {\it not} local quantum
field theories \rsdecoupled.  This non-locality distinguishes them
{}from the $(2,0)$ field theories which also have string-like
excitations, but appear to be ordinary local quantum field theories.

It is natural to ask whether we can find more theories by
compactifying more transverse directions, or by considering
five-branes on more general geometries than flat non-compact
transverse spaces.  In section two, we study the NS five-branes in IIA
and IIB with a transverse circle.  This circle compactifies some of
the directions in the moduli space of vacua of these theories.
However, the size of these directions has a factor of $1 \over g_s$
($g_s$ is the type IIA or type IIB string coupling) relative to the
naive expectation.  This factor has significant consequences.  First,
to decouple the physics in the bulk in order to find a complete theory
on the brane, we have to set the string coupling to zero.  This leads
to the decompactification of these directions.  Therefore, the brane
theories with vanishing string coupling in these cases are the same as
those in \rsdecoupled.  Second, it shows that unlike D-branes \rpol,
which can serve as probes \refs{\doumoo,\rprobes}, the NS five-branes
do not probe the background transverse space at zero string coupling.

In section three, we focus on the NS five-branes in type IIA string
theory with a transverse circle of radius $R_A$ at arbitrary string
coupling. We consider the matrix model \rBFSS\ description of this
configuration.  It is given by a 2+1 dimensional field theory similar
to that of \rbd.  In the limit $R_A \rightarrow \infty$, the theory
becomes a 1+1 dimensional theory describing the NS five branes in
type IIA theory.  For $R_A \rightarrow 0$ another 1+1 dimensional
theory appears, which describes an $A_{k-1}$ singularity in IIB theory.
Our 2+1 dimensional theory interpolates between these two limits.
For $g_s^A=0$, its Higgs branch decouples {}from its Coulomb branch and
gives the matrix model description of the $(2,0)$ string theory
\refs{\rquantumfive,\rstringfive}.

\newsec{String Theory NS five-branes on $\IR^9\times S^1$}

\subsec{Type IIA five-branes on $\IR^9\times S^1$}

We start by considering M theory on $\IR^9 \times T^2$ with $k$
five-branes at points on the torus $T^2$.  The $T^2$ is defined by its
complex structure $\tau$ and its volume. For simplicity, let us take
the torus to be rectangular with radii $ R_1$ and $R_2$. This choice
will not affect any of the following discussion in a significant
way. The moduli space for the five-branes is then
\eqn\modulispace{ \CM={(\BR^3 \times T^2)^k \over {\bf S}_k}. }
Using the argument leading to \tzsr\ we find that the radii of the
$T^2$ factors are
\eqn\periods{ \eqalign{ &P_1 = M_{pl}^3 R_1 \cr
                        &P_2 = M_{pl}^3 R_2. }}
M theory on a two-torus is equivalent to the type IIB string \rmIIB\
on a circle of radius
\eqn\rbdef{R_B = {1\over M_{pl}^3 R_1 R_2} = {1\over M_s^2 R_2},}
with string coupling
\eqn\twobc{g_s^B = {R_1\over R_2}.}
This result can be obtained by going {}from M theory to type IIA on
$R_1$.  Then T duality on $R_2$ maps us to the IIB theory on a circle
with radius \rbdef\ and coupling \twobc.  The NS five-branes in the
IIA theory, which are located at points on the transverse circle
$R_2$, are mapped under the T duality into Kaluza-Klein monopoles
\rTduality.  Therefore, our theory of $k$ M theory five-branes on
$T^2$ is a theory of $k$ Kaluza-Klein monopoles in type IIB string
theory.

Let us review some basic facts about Kaluza-Klein monopoles.  The
monopole solution is constructed by taking flat space tensored with
the four dimensional multi-Taub-NUT metric \rmultitn.  In the case of
string theory, this construction gives a five-brane, while in M
theory, we obtain a six-brane \rsixbrane. The non-trivial metric on $
\IR^3 \times S^1$ is
\eqn\multitn{ ds^2 = V(x)\, d\vec{x}^2 + V(x)^{-1} (d\theta + \vec{A}
\cdot d\vec{x})^2,}
where $\vec{x}$ is three dimensional and $\vec{A}$ is related to $V$
by
\eqn\varl{\grad {} V = \grad {} \times \vec{A}.}
The scalar function $V$ depends on a single free parameter $r$:
\eqn\definer{ V = 1  + r \sum_{i=1}^k {1\over |\vec{x}-\vec{x}^i|}. }
The positions of the $k$ branes are specified by the $ \vec{x}^i $
and the angular variable $\theta$ has a period proportional to $r$.
The parameter $r$ sets the scale of the solution, and can be
rescaled by rescaling $\vec x$ and $\theta$.  It corresponds to the
size of the circle $S^1$ in the limit $ |x| \r \infty$.  In our
problem $r=R_B$.  When all the branes are separated the space is
smooth.  For $k>1$ coalescing branes the multi-Taub-NUT has an
$A_{k-1}$ singularity at the position of the branes.  In the limit $r
\r \infty$ the circle which is coordinatized by $\theta$
decompactifies everywhere except at the positions of the branes and
the space becomes $\IR^4/Z_k$.  Finally, we should mention that the
multi-Taub-NUT has a number of non-trivial two-cycles.  Some of these
cycles collapse when the $\vec x^i$ coalesce; note that there is a
non-trivial two-cycle even for $k=1$ \rtwoform.  

For recent discussions of Kaluza-Klein monopoles in M
theory and string theory see
\refs{\rashoke\rgeneralsixbrane\kutetal\rmoore\rbrunner -\rgilad}.

It is useful to re-express the relations \periods\ in terms of the
string scale and string couplings
\eqn\newperiods{\eqalign{
P_1 &=  M_{s}^2  \cr
P_2 &= { M_s^2 \over g_s^B} = {M_s^3 R_2 \over g_s^A}. }}
The key feature is that $P_2$ always contains a factor of $1/g_s$
whether expressed in terms of the type IIA or type IIB string
coupling.

There is a simple reason for these factors of $1 \over g_s$.  The
collective coordinates of each NS five-brane in IIA, or Kaluza-Klein
monopole in IIB, are a two form and five scalars (for a recent
discussion, see \rashoke).  The two form and one of the scalars
$\Phi_1$ arise {}from the RR sector -- for the type IIB Kaluza-Klein
monopole they arise {}from the RR four form and the RR two form
reduced on the non-trivial two-cycle.  The other four scalars are
NS-NS fields -- they correspond to the three deformations of the
metric, and a compact deformation $\Phi_2$ of the NS-NS two form.  The
natural normalization of these fields is with a factor of $1 \over
g_s^2$ in front of the kinetic terms for the NS-NS fields, but not in
front of the kinetic terms for the RR fields.  In order to keep the
$(2,0)$ supersymmetry on the five-brane manifest, we rescale the NS-NS
scalars to have no $1 \over g_s^2$ in their kinetic terms.  This leads
to the crucial factor of $1 \over g_s$ in $P_2$.  

This situation should be contrasted with that of D-branes.  There, all
the collective coordinates appear {}from open strings.  Both the gauge
fields and the scalars have the same normalization, $1 \over g_s$, in
their kinetic terms and therefore no rescaling is necessary.
Therefore, these scalars ``see'' the underlying geometry, and D-branes
can be used as probes.  On the other hand, the NS five-branes and the
Kaluza-Klein monopoles are not good probes. In particular, for finite
$R_2$ the value of $P_2$ diverges as the string coupling goes to zero.

We now want to decouple the bulk physics to obtain a complete theory.
This can be accomplished only if $g_s^A=g_s^B=0$.  It is clear {}from
\newperiods\ that in this case $P_2=\infty$.  A more careful analysis
immediately shows that this conclusion cannot be avoided by taking
various limits of $R_1$ and/or $R_2$.  For example, if we take $g_s^A,
R_2 \r 0$, while holding $R_2 \over g_s^A$ fixed, $P_2$ is finite.
However, since $R_2\r 0$ this theory is better thought of as the type
IIB theory in $\IR^{10}$ with a finite coupling, and the bulk physics
no longer decouples.

The spacetime geometry in this limit depends on $R_2$. When $R_2 \r
0$ the type IIB Kaluza-Klein monopoles are the better description,
while for $R_2 \r \infty$ the type IIA NS five-branes are the right
description. However, the decoupled physics on the brane is actually
independent of $R_2$ since $P_2$ has gone to infinity.  We will find
further evidence favoring the uniqueness of this decoupling limit
{}from the matrix theory description of this configuration, discussed
in the following section.

Our analysis leads us to a description of the decoupled physics on
$A_{k-1}$ singularities in free type IIB string theory.  It is given
by the same ``non-critical string theory'' as the $k$ NS five-branes
in type IIA theory.  This fact is known for the low energy $(2,0)$
field theories, and here we recover it for the $(2,0)$ string theory.
The $(2,0)$ string theory has two kinds of strings: those which exist
in the $(2,0)$ field theory, whose tension vanishes at the
singularities in the moduli space, and other strings with tension
$M_s^2$.  In the IIA description both kinds of strings are membranes
stretching between five-branes and wrapping the compact direction.  In
the type IIB $A_{k-1}$ theory, strings of the first kind are
associated with IIB three-branes which wrap collapsing two-cycles.
Strings of the second kind are bound states at threshold of strings
{}from the bulk with the Kaluza-Klein monopoles.

Essentially the same scaling analysis applies when more transverse
circles are present. The extra factor of $1 \over g_s$ rescales the
metric, and decompactifies the transverse space in the $g_s \r 0$
limit.  Furthermore, we can consider type IIA five-branes with an
arbitrary smooth transverse metric.  It seems that a similar factor of
$1\over g_s$ would make the general target space geometry as
``probed'' by the five-branes flat and non-compact in the decoupling
limit.  The case of five-branes at a singularity will be discussed in
\kin.

\subsec{Type IIB five-branes on $\IR^9\times S^1$}

We now consider the case of $k$ type IIB five-branes whose world
volume theory has $(1,1)$ supersymmetry.  We start with M theory
compactified on a torus $T^2$ with radii $R_1$ and $R_2$.  The type
IIB five-branes arise as M theory Kaluza-Klein monopoles associated
with one of the cycles which wrap the other cycle.  For example, let
us go {}from M theory to type IIA by reducing on $R_1$, and consider
Kaluza-Klein monopoles associated with $R_1$, so that their $r$
parameter is $R_1$.  These solitons are D6-branes in the type IIA
string theory.  T duality on $R_2$ maps us to the type IIB theory, and
the Kaluza-Klein monopoles become D5-branes at points on a circle of
radius $R_B={1 \over M_{pl}^3 R_1 R_2}$.  S-duality converts them to NS
five-branes at points on the circle.  Instead, we can start with
Kaluza-Klein monopoles in the IIA theory associated with $R_2$, so
their $r$ parameter is $R_2$, and T dualize $R_2$ to find the NS
five-branes of IIB at points on a circle of radius $R_B$.

The low energy theory on the NS five-branes is a $U(k)$ gauge theory
with gauge coupling $1 \over M_s^2$ \rsdecoupled.  When $R_B$ is
finite, the moduli space of vacua of this theory is
\eqn\modoneone{{(\IR^3 \times S^1)^k \over {\bf S}_k}.}
The radius $P$ of the $S^1$ factors is easy to determine, e.g.\
by starting with the wrapped D6-brane description in the previous
paragraph, and performing the duality transformations.  We find:
\eqn\IIBperiod{P ={M_s^2 R_B \over g_s^B}= {M_s\over g_s^A}. } 
As in the previous subsection, we see that $P$ has a factor of $1
\over g_s$ relative to the naive result.  As we said there, this is
unlike the case of D-branes.  This factor of $1\over g_s$ can be
explained as in that case.  The gauge fields are RR fields, which for
the Kaluza-Klein monopole arise {}from the reduction of the three form
on the non-trivial two-cycle, while the four scalars are NS-NS fields.
Rescaling the NS-NS scalars to have the same kinetic terms as the one
forms leads to \IIBperiod.

The factor of $1 \over g_s$ also has consequences for the possible
decoupling limits. Decoupling requires taking $g_s^A, g_s^B \r
0$. Once again, the period for the scalar decompactifies, and we are
driven back to the theory of parallel type IIB five-branes in
$\IR^{10}$. Also, in analogous fashion to the case with $(2,0)$
supersymmetry, there is a parameter $R_B$, which changes the spacetime
description, but does not alter the decoupled physics.

\newsec{A  Matrix Definition of M theory Five-Branes on Compact Spaces}

A matrix model \rBFSS\ for the M theory five-brane on $\IR^9\times
T^2$ follows naturally by extending the quantum mechanics describing
the longitudinal five-brane \rbd\ to $k>1 $ five-branes, and to a 2+1
dimensional field theory with eight supersymmetries.  For related
discussions see \refs{\doumoo,\rfischler}.  The theory has a $U(N)$
gauge symmetry where $N$ is the number of zero-branes used to probe
the longitudinal five-brane. The coupling to $k$ parallel five-branes
is represented by $k$ hypermultiplets in the fundamental of the gauge
group. There is also an adjoint hypermultiplet, which encodes motion
of the zero-branes within the longitudinal five-brane.

The interaction of the five-branes with spacetime is encoded in the
dynamics on the Coulomb branch of the theory. For $k>1$, there is also
a Higgs branch in the model that corresponds to physics localized
within the brane. Points on the Higgs branch essentially describe the
dynamics of zero-branes, which have fattened to instantons within the
five-branes \rbraneswith. 

The parameters of the matrix theory are determined in terms of the
radius of the longitudinal direction $R$, and the two radii $R_1$ and
$R_2$ of the compact part of spacetime $T^2 \times\IR^7$.  In terms
of these parameters \refs{\rBFSS, \rwati\rSS-\rBS}, the
2+1 dimensional theory is on a compact space with radii
\eqn\ymdim{ \Sigma_i = {1 \over M_{pl}^3 R_i R},}
and the Yang-Mills gauge-coupling is
\eqn\ymcoupling{ g^2_{YM} = {R \over R_1 R_2} =  R^3
M_{pl}^6 \S_1 \S_2. }
It is convenient to express the dimensions of the torus in terms of
the string scale and string coupling
\eqn\newymdim{ \eqalign{ \S_1 &= {1 \over M_s^2 R} \cr
                         \S_2 &= {1\over M_s^2 R} g_s^B.}}

Consider first the 2+1 dimensional theory on $\IR^3$.  Both the Higgs
and the Coulomb branches are described by hyper\kh manifolds. A
non-renormalization theorem guarantees that the Higgs branch is immune
to quantum corrections \raps.  In terms of the fields in the
Lagrangian, it provides the ADHM hyper\kh quotient construction of the
moduli space of $N$ instantons in $SU(k)$ gauge theory in four
dimensions.  The Coulomb branch of the theory for $N=1$ was analyzed
in \refs{\rsprobes,\rswthree}.  Its metric is a Taub-NUT metric.  For
higher $N$ the metric appears to be a symmetric product of Taub-NUT
metrics.  The Coulomb and Higgs branches touch at a singular point
where the theory flows to a non-trivial interacting three dimensional
fixed point.  The infra-red limit is the same as taking the
dimensionful coupling constant $g^2_{YM} \rightarrow
\infty$.  It is a property of this fixed point that the Higgs branch and
the Coulomb branch both emanate {}from it.  Therefore, these two
branches are not decoupled here.

Now we consider the theory with finite $\S_{1,2}$. There can be Wilson
lines on $T^2$, but we will ignore them.  We are going to explore this
theory for fixed $\S_1$ as a function of $\S_2 \ll \S_1$ and $
g^2_{YM} \gg 1/\S_1 $.  For $N=1$, this problem was analyzed in
\rthroat. The relevant dimensionless quantity which controls the
dynamics is
\eqn\parameter{ \gamma = g^2_{YM} \S_2 = {1 \over (R_2 M_s)^2}.}  
Consider first the limit $\gamma \gg 1$.  At energies larger than $ 1/
\S_2$ the theory is three dimensional and its Coulomb branch becomes a
symmetric product of Taub-NUT spaces. At energies of order $1 \over
\S_2$ the theory becomes two dimensional.  The two dimensional sigma
model based on the Taub-NUT metric is conformally invariant, and
therefore this metric does not change as we flow to the infra-red.  In
the opposite limit, $\gamma \ll 1$, the theory becomes
two dimensional at the scale $1 \over \S_2$ before the gauge
interactions become strong.  Therefore, here the dynamics is that of
the two dimensional gauge theory.  The result of this dynamics is a
metric with an infinite tube \rchs.  For $N=1$, the explicit answer
which interpolates between the $1 / |x|$ behavior for $\gamma \gg 1$
and the $1 / |x|^2$ behavior for $\gamma \ll 1$ was found in \rthroat.

These results are consistent with the spacetime picture. The parameter
$\gamma=1/(R_2M_s)^2$ interpolates between the two and three
dimensional theories, which are appropriate to type IIA and type IIB,
respectively.  For $R_2 \gg 1/M_s$ the metric we expect is the tube
metric of the NS five-brane of the IIA theory \rchs, while for $R_2
\ll 1/M_s$ we expect the Taub-NUT metric of the Kaluza-Klein monopoles
in type IIB theory.  It is satisfying to see how the matrix model
reproduces these answers.

We can now consider the decoupling limit described in the previous
section in the context of this matrix model. In this limit $\S_2 \r
0$, $g^2_{YM} \r \infty$ while $\S_1$ and $\gamma$ are fixed.  The 2+1
dimensional theory becomes 1+1 dimensional.  Now we can use the
arguments of \refs{\rwtensor, \rquantumfive, \rstringfive} to argue
for the decoupling of the Higgs branch and the Coulomb branch in this
limit.  For this decoupling it is crucial that we consider the two
limits $\S_2 \r 0$ and $g^2_{YM} \r \infty$.  In particular, without
the $\S_2 \r 0$ limit, the theory is 2+1 dimensional where no such
decoupling happens.

It is interesting to examine the $R_2$ or the $\gamma$ dependence in
this limit.  The physics of the Higgs branch is independent of these
parameters.  This follows {}from the non-renormalization theorem
mentioned above, as well as {}from the fact that the Higgs branch
metric is independent of $\S_2$ \rgms. This independence is in accord
with our statements in the previous section about the decoupled
physics on the five-brane being independent of $R_2$.  On the other
hand, as mentioned above, the Coulomb branch depends on $R_2$
corresponding to the fact that the spacetime metric depends on $R_2$.

There is a subtlety that is worth mentioning. In the case without
the longitudinal five-brane, the 2+1 dimensional theory has sixteen
supersymmetries. This theory has an interacting fixed-point with
$Spin(8)$ global symmetry \refs{\rSS, \rBS, \rreview}. In this case,
there are two inequivalent limits in which the field theory becomes
1+1 dimensional. The first is dimensional reduction to Yang-Mills in
two dimensions. Flow to the infra-red gives an orbifold conformal
field theory which describes the type IIA string theory \refs{\rmotl,
\rBS, \rDVVstring}. The second is obtained by first flowing to the 
2+1 dimensional fixed point, and then reducing to 1+1 dimensions. In
this case, the resulting 1+1 dimensional conformal field theory
describes the type IIB string. The difference is essentially in the
way that the extra spacetime dimension is acquired. In the first case
by dimensional reduction, while in the second, by dualizing the
gauge-field in the Abelian case, or flowing to the interacting fixed
point for the non-Abelian case. By contrast, in the situation with the
longitudinal five-brane, we are interested in the Higgs branch of the
theory, and then the two limits commute. There is only one decoupled
1+1 dimensional conformal field theory: the theory which describes
parallel type IIA five-branes.

\bigbreak\bigskip\bigskip\centerline{{\bf Acknowledgements}}\nobreak

It is a pleasure to thank O. Aharony, K. Intriligator and E. Witten
for helpful discussions. The work of N.S. is supported in part by
\#DE-FG02-90ER40542; that of S.S. by NSF grant DMS--9627351.

\listrefs

\end